\documentclass[a4paper,11pt]{article}
\pdfoutput=1 

\usepackage{jinstpub} 

\usepackage{array}
\usepackage{graphicx}      
\usepackage{tabularx}          
\usepackage[utf8]{inputenc}
\usepackage{array}
\usepackage{makecell}

\title{\boldmath Precise Tracking of Cosmic Muons using Time-over-Threshold Property of NINO ASICs}

\author[a,b,1]{Sridhar Tripathy\note{Corresponding author}}
\author[a,b]{Subhendu Das}
\author[a,b]{Jaydeep Datta}
\author[a,b]{Nayana Majumdar}
\author[a,b]{and Supratik Mukhopadhyay}

\affiliation[a]{Saha Institute of Nuclear Physics,\\AF Block, Sector 1, Salt Lake, Kolkata 700064, India}
\affiliation[b]{Homi Bhabha National Institute, \\Training School Complex, Anushaktinagar, Mumbai 400094, India}

\emailAdd{sridhar.tripathy@saha.ac.in}

\abstract{ This work reports a cost-effective, simple front-end readout and DAQ of a RPC-based muon scattering tomography system under construction. The Time-over-threshold
property of NINO ASICs has been exploited to achieve precise tracking by extraction of position information. The use of a low-cost FPGA for event selection and TOT measurement has been demonstrated for a prototype RPC which has been triggered by a cosmic ray hodoscope of three plastic scintillators. The relation between input pulse width and TOT has been established by analyzing the waveforms of the RPC signals. The TOT profile at different working voltages has been obtained by the FPGA and it has been fitted with the Gaussian distribution. Its standard deviation represents an estimate of the tracking precision. }


\arxivnumber{ 2004.14984 } 

\proceeding{XV$^{\text{th}}$ Workshop on Resistive Plate Chambers and Related Detectors\\
  10-14 February, 2020\\
  Rome, Italy}

\begin{document}
\maketitle
\flushbottom

\section{Introduction}
\label{sec:intro}
\begin{figure}[h]
	\centering
	\includegraphics[trim = 0 0 0 0, clip, angle = 0, width=0.36\textwidth,height=0.35\textwidth]{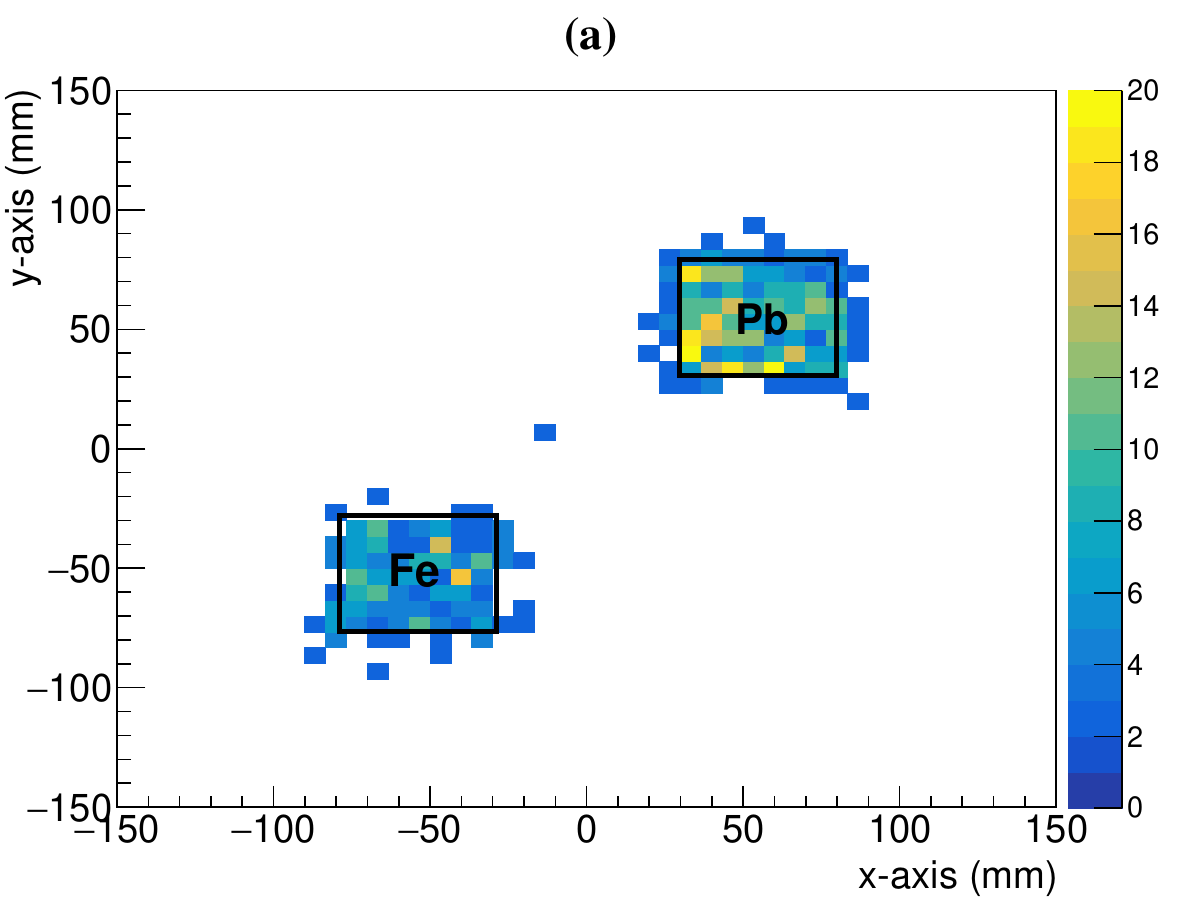}
	\includegraphics[trim = 0 0 0 0, clip, angle = 0, width=0.36\textwidth,height=0.35\textwidth]{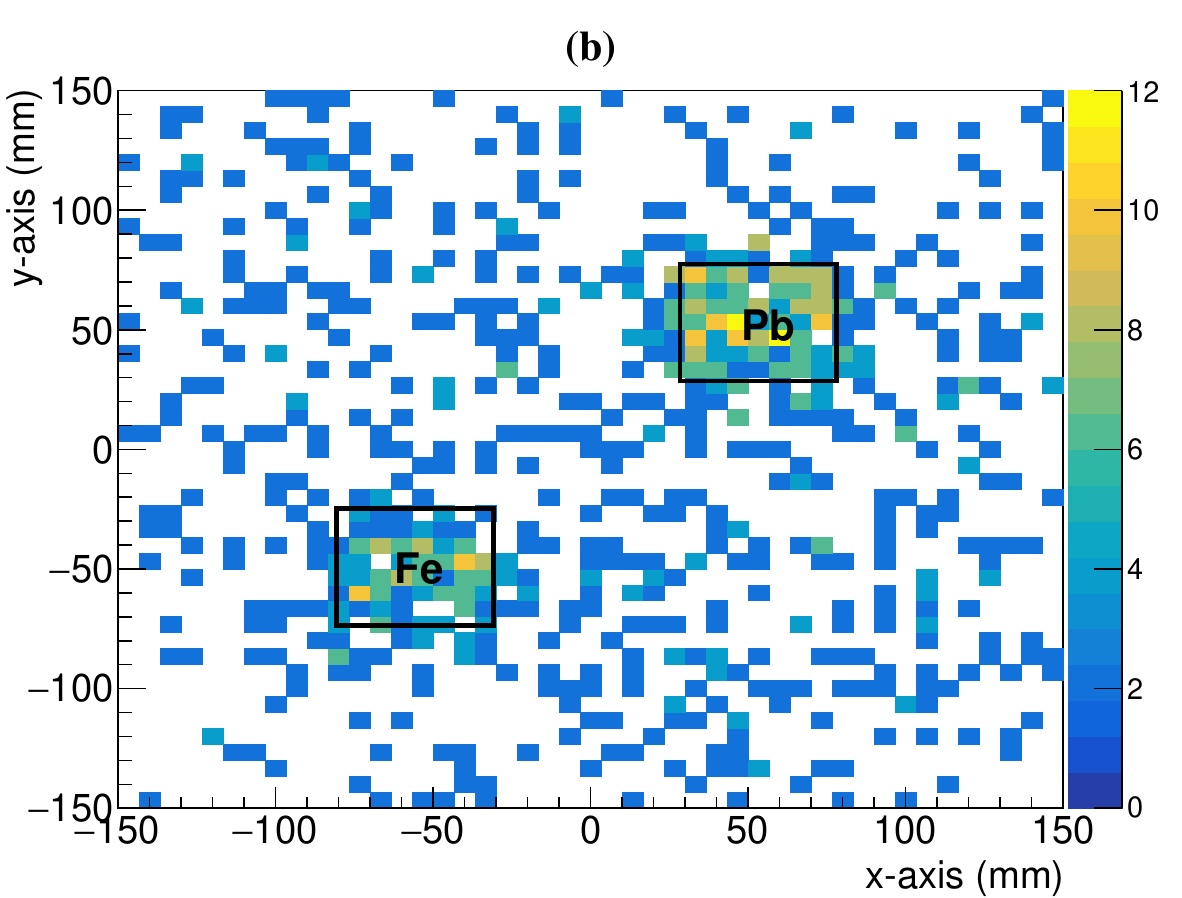}
	\caption{\label{resolution} The reconstructed images for two different spatial resolutions: (a) 200 $\mu$m, (b) 1 mm}
\end{figure}

Muon Scattering Tomography (MST) exploits the deviation of cosmic muons from their path due to their interaction with atomic nuclei and electron of the target material to probe its physical properties. Gaseous detectors have proven to be  versatile means for tracking the cosmic muons \cite{baesso},\cite{tsinghua},\cite{moris} and therefore are used widely in MST applications. The RPC is one of the popular gaseous detectors which is found suitable for this purpose due to its simple design, ease of construction, cost-effectiveness, large-scale producibility, along with very good temporal, spatial resolutions, and detection efficiency. The spatial resolution of the RPC determines the precision in tracking and hence the quality of image formation. It can be followed from an example of the images of two objects produced in a MST setup for two different spatial resolutions as shown in figure~\ref{resolution}. An elaborate study on the effect of spatial resolution on image formation is available in a previous simulation work \cite{arxv}. Controllable operational parameters such as, applied electric field, gas mixture etc. are a few factors that influence the detector spatial resolution. But most importantly, it is determined by the read-out granularity as the position information is extracted from the distribution of charge induced on the read-out strips. Finer strips can predict the center of a charge distribution with better resolution which helps to achieve pin-pointed position information. For a read-out with finer strips a larger number of channels are required and hence the cost of electronics increases significantly. A multi-channel read-out ASIC which also serves the purpose of an ADC, can be a possible solution for precise tracking as well as inexpensive electronics.

In this work, the authors have suggested a reliable option for extracting charge information based on the Time-Over-Threshold~(TOT) property of NINO~\cite{NINO}, which is a low power, ultra-fast front-end ASIC designed for ALICE-TOF detectors~\cite{tifrnino},~\cite{alice}. By virtue of its TOT property, the NINO ASIC can represent charge in terms of time which can be obtained from the width of its output pulse~\cite{jj}, \cite{flextot}. The DAQ system has been designed with an FPGA for time measurement using a high-frequency clock. The objective of this work is to study the performance of the proposed scheme of read-out and DAQ for extracting position information. This has been investigated using a prototype RPC detector and the results have demonstrated that this can be useful for building read-out electronics of a RPC-based MST set-up.
 
\section{Components of the Read-out \& DAQ}
\label{sec:DAQ}
The proposed scheme comprises of two components: NINO used for front-end read-out and FPGA in the back-end for acquiring the data. These two have been briefly described below.
\subsection{NINO in Front-end}
The NINO is a 8-channel LVDS input/output discriminator, pre-amplifier ASIC. It has been designed with the TOT property, i.e. the output pulse width of NINO is related to the input detector pulse width above a pre-set threshold. For a given detector and working conditions, it is assumed that the TOT is related to the amplitude of the detector signal which is in turn a measure of the induced charge~\cite{gonnela},\cite{becker}. Therefore, the charge deposit can be related to time which can be digitized using a TDC. 

\subsection{FPGA in back-end}
\label{sec:FPGA}
\begin{figure}[h]
	\centering
	\includegraphics[trim = 0 0 0 0, clip, angle = 0, width=0.4\textwidth]{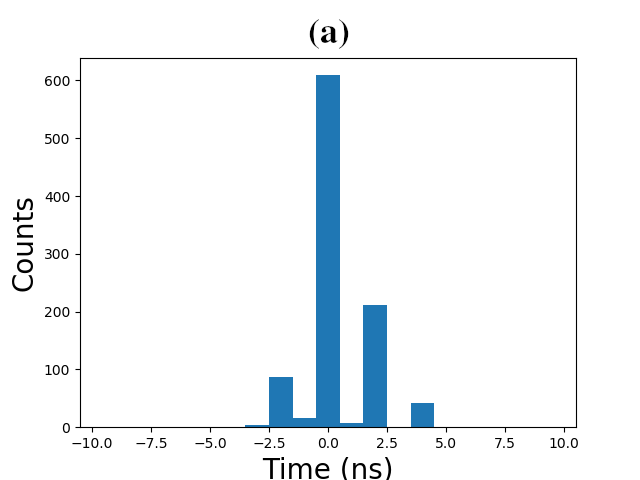}
	\includegraphics[trim = 0 50 0 20, clip, angle = 0,width=0.4\textwidth]{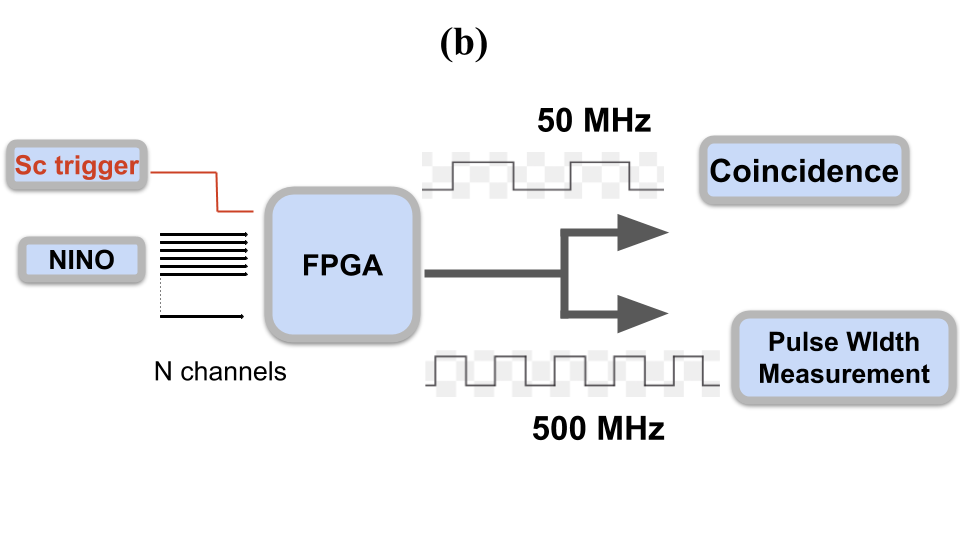}
	\caption{\label{FPGADAQ} (a) Difference of pulse widths measured by the oscilloscope and the FPGA, adding an offset of 2 ns, (b) Schematic diagram of the operation inside the FPGA}
\end{figure}

Taking the advantage of the low rate of cosmic muons, a low-end FPGA has been used in the present work to design a TDC of coarse resolution.
A developer board based on ALTERA MAX-10 FPGA has been used for pulse width measurement~\cite{fpga}. It has been programmed using the Intel Quartus Prime platform. The TOT output of NINO has been measured by a 500 MHz clock generated using Phase Locked Loop (PLL) mechanism applied on the 50 MHz on-board clock. Therefore, a resolution of 2 ns has been achieved in this scheme. It has been tested by measuring the pulse width of a known 60 ns pulse by both oscilloscope and FPGA. The difference in their measurement has been shown in figure~\ref{FPGADAQ}~(a) with an offset of 2 ns. The following logical operation has been programmed in the FPGA for collecting data from the RPC. The TOT values for seven read-out strips (center strip, three on the right, three on the left) are temporarily stored in a memory array. These data are serially transmitted to the computer via UART protocol when a coincidence of three scintillators and the central strip of RPC is found. The schematic workflow of the FPGA has been illustrated in figure~\ref{FPGADAQ} (b). 

\section{Experimental Setup}
\label{sec:Exp}
\begin{figure}[h]
	\centering
	\includegraphics[trim = 0 0 0 0, clip, angle = 0,width=0.39\textwidth]{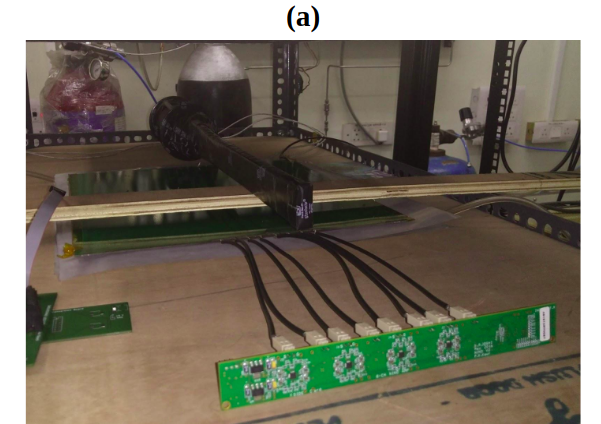}
	\includegraphics[trim = 0 0 0 0, clip, angle = 0,width=0.45\textwidth,height=0.21\textwidth]{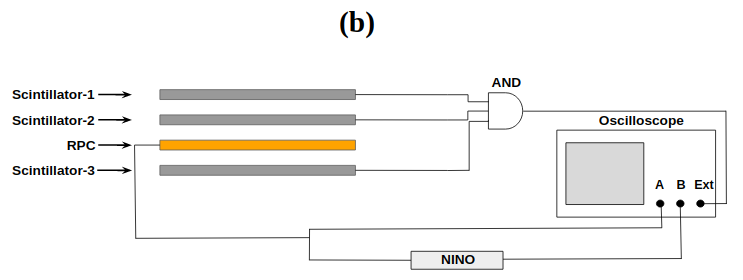}
	\caption{\label{expset} (a) Image of the experimental setup, (b) Schematic diagram of the setup for the measurement of input pulse width (connected in channel A of oscilloscope) and TOT output of NINO (connected in channel B of oscilloscope).}
\end{figure}

A prototype RPC detector has been used for the testing of proposed scheme of read-out electronics. The dimension of the RPC is 30~cm~$\times$~30~cm and it is made of Bakelite. A mixture of R134a (95\%) and iso-butane (5\%) gases has been used. The read-out strips are made of copper and have width 1 cm, pitch 1.2 cm and length 30 cm. The RPC has been placed in a vertical cosmic ray hodoscope of three plastic scintillators with an overlapped area that matches the width of a single (central) read-out strip of the RPC. The finger scintillator among the three, has been kept as close as possible to the RPC to trigger only a single strip. However, waveforms from three strips from either sides have been studied to obtain the charge spread of a signal. An image of the experimental setup has been shown in figure~\ref{expset}~(a). Two key aspects have been investigated in this experiment. Firstly, the raw pulse of RPC and NINO TOT outputs have been compared by analyzing their waveforms using Tektronix MSO 4104-b oscilloscope. This process has been illustrated in figure~\ref{expset}~(b). Secondly, the profile of induced charge on the strips has been obtained using the TOT measured by the FPGA. 

\begin{figure}[h]
	\centering
	\includegraphics[trim = 0 0 0 0, clip, angle = 0,width=0.48\textwidth,height=0.30\textwidth]{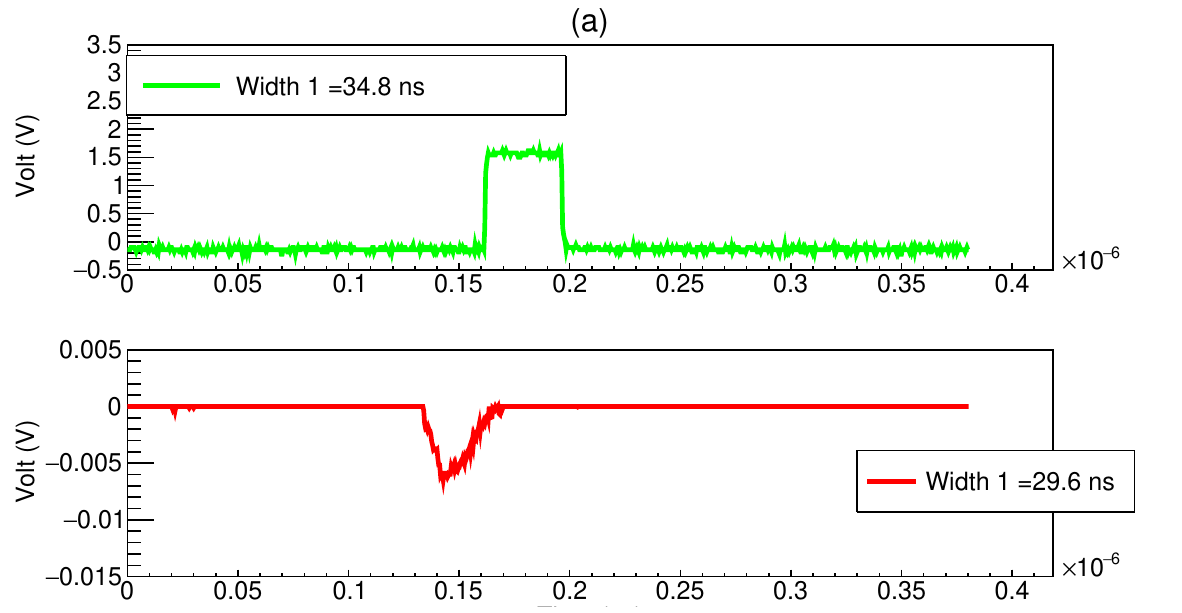}
	\includegraphics[trim = 0 0 0 0, clip, angle = 0,width=0.48\textwidth,height=0.30\textwidth]{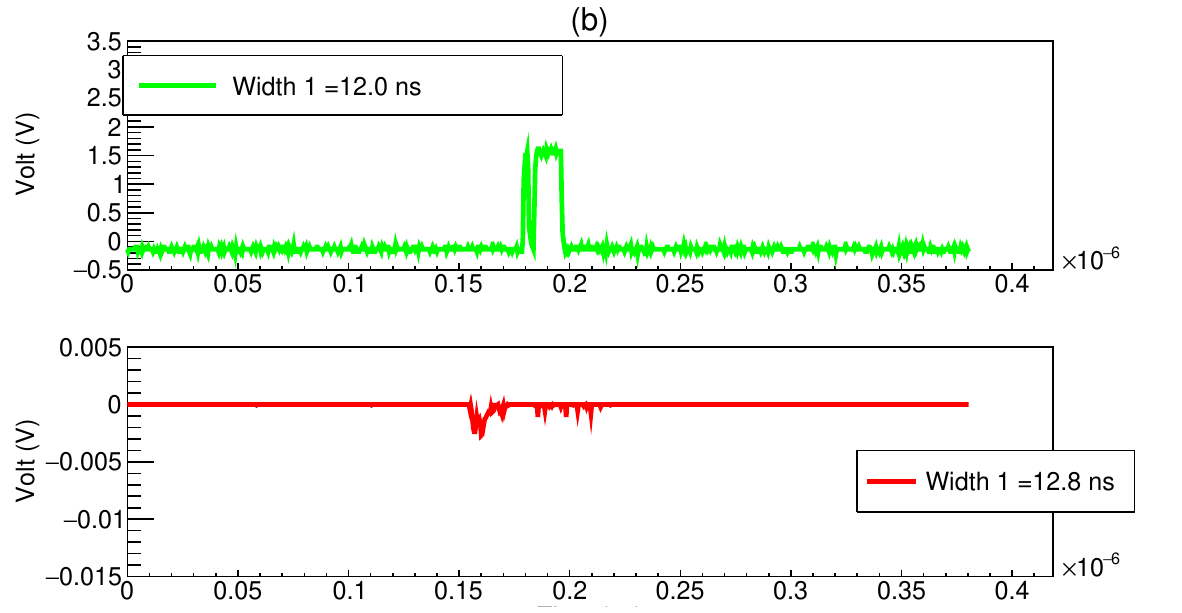}
	\caption{\label{107} Comparison of RPC pulse and NINO TOT output for (a) large pulse and (b) a smaller pulse.}
\end{figure}

In figure~\ref{107}, two typical RPC pulses at a given voltage and their corresponding NINO outputs as obtained with the oscilloscope, have been shown. The following criteria have been used for selecting the RPC pulses for comparison to the NINO output. The RPC pulse generally arrives around the same position on time scale. Therefore, the first 200 ns window of the pulse has been considered as background electronic noise. The average amplitude of the noise window has been considered as a threshold for a valid pulse. It is much higher than the maximum threshold that can be set by NINO board (100 fc). Those pulses which last at least 8 ns above the threshold are taken as a valid pulse. If a signal has multiple rises and falls along with the existence of several valid pulses, the signal is considered to have multiple pulses. This indicates the existence of streamers.
\begin{figure}[htb]
	\centering
	\includegraphics[trim = 0 0 0 0, clip, angle = 0,width=0.4\textwidth]{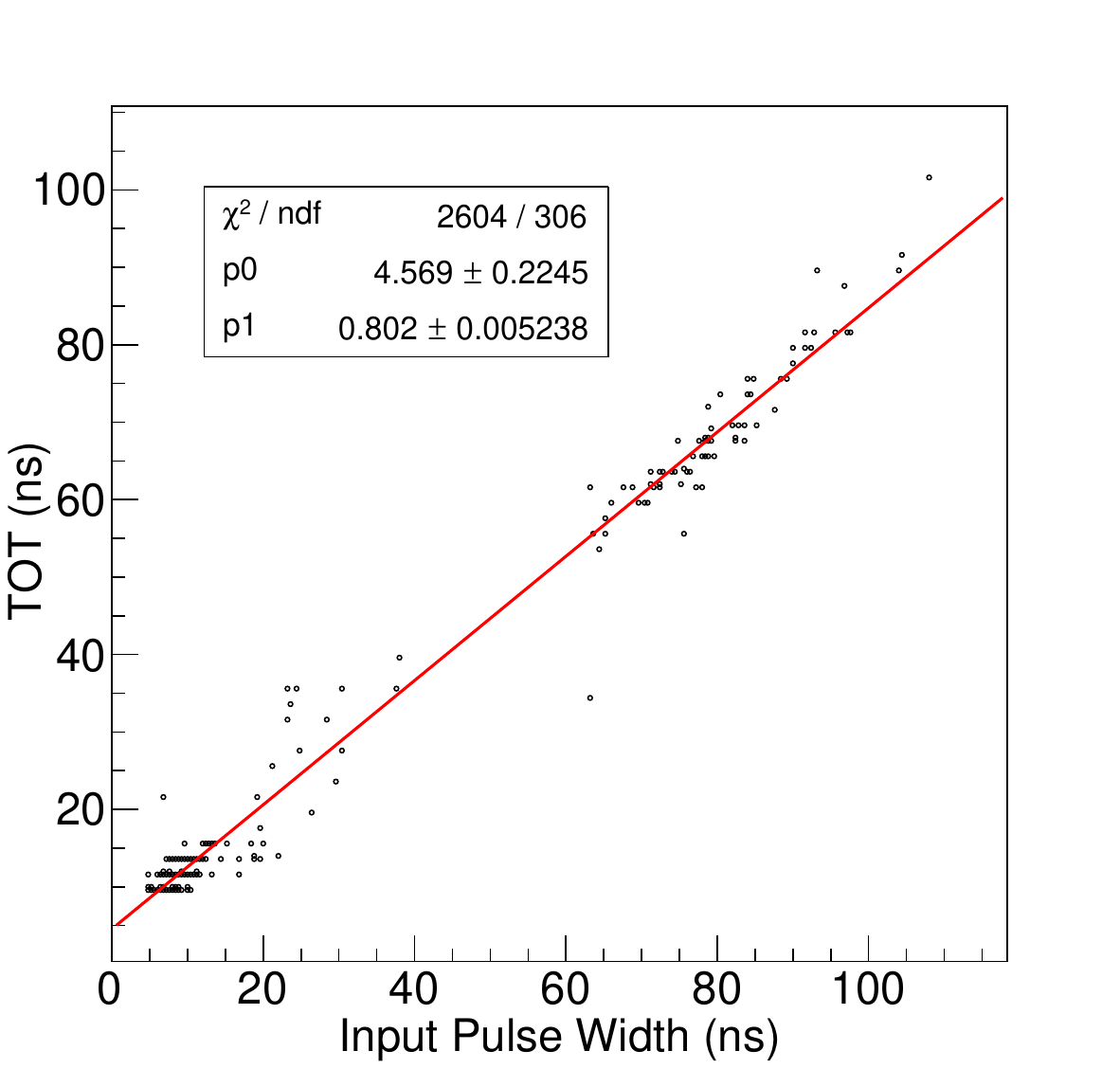}
	\caption{\label{widthTOT} Variation of input pulse width with output TOT}
\end{figure}
\begin{figure}[h]
	\centering
	\includegraphics[trim = 0 0 0 0, clip, angle = 0,width=0.9\textwidth,height =0.4\textwidth ]{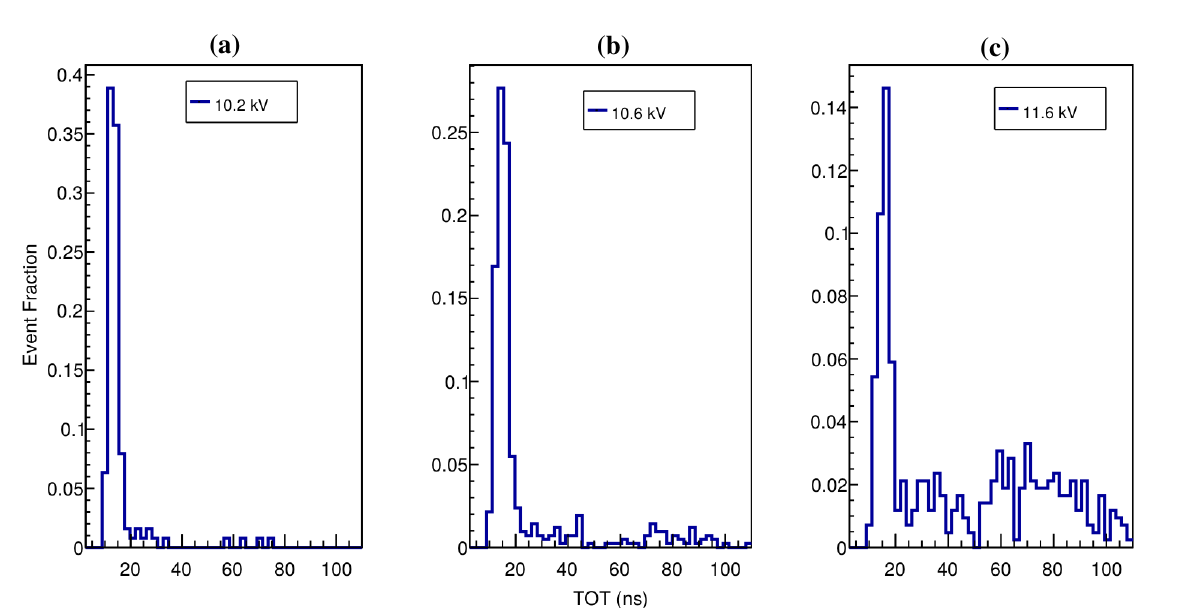}
	\caption{\label{DiffVolt}TOT of the central strip for working voltages:~(a) 10.2 kV, (b) 10.6 kV and (c) 11.6 kV}
\end{figure}
\begin{figure}[h]
	\centering
	\includegraphics[trim = 0 0 0 0, clip, angle = 0,width=0.9\textwidth,height =0.4\textwidth]{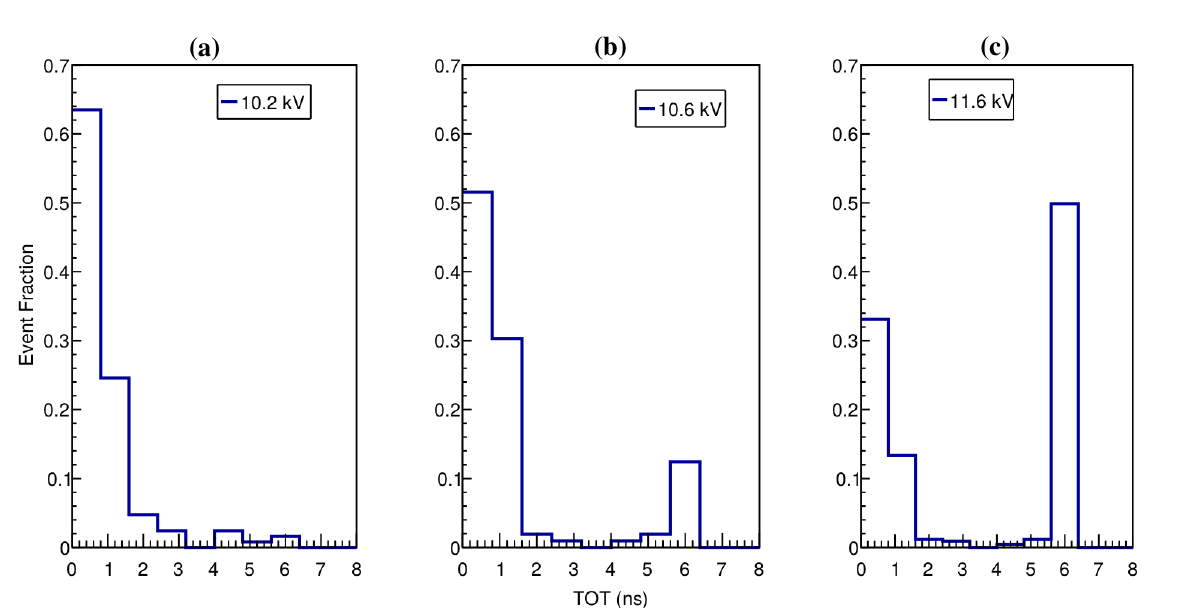}
	\caption{\label{StripM} The strip multiplicity for working voltages:~(a) 10.2 kV, (b) 10.6 kV and (c) 11.6 kV}
\end{figure}
\begin{figure}[h]

	\centering
	\includegraphics[trim = 0 0 0 0, clip, angle = 0,width=0.4\textwidth]{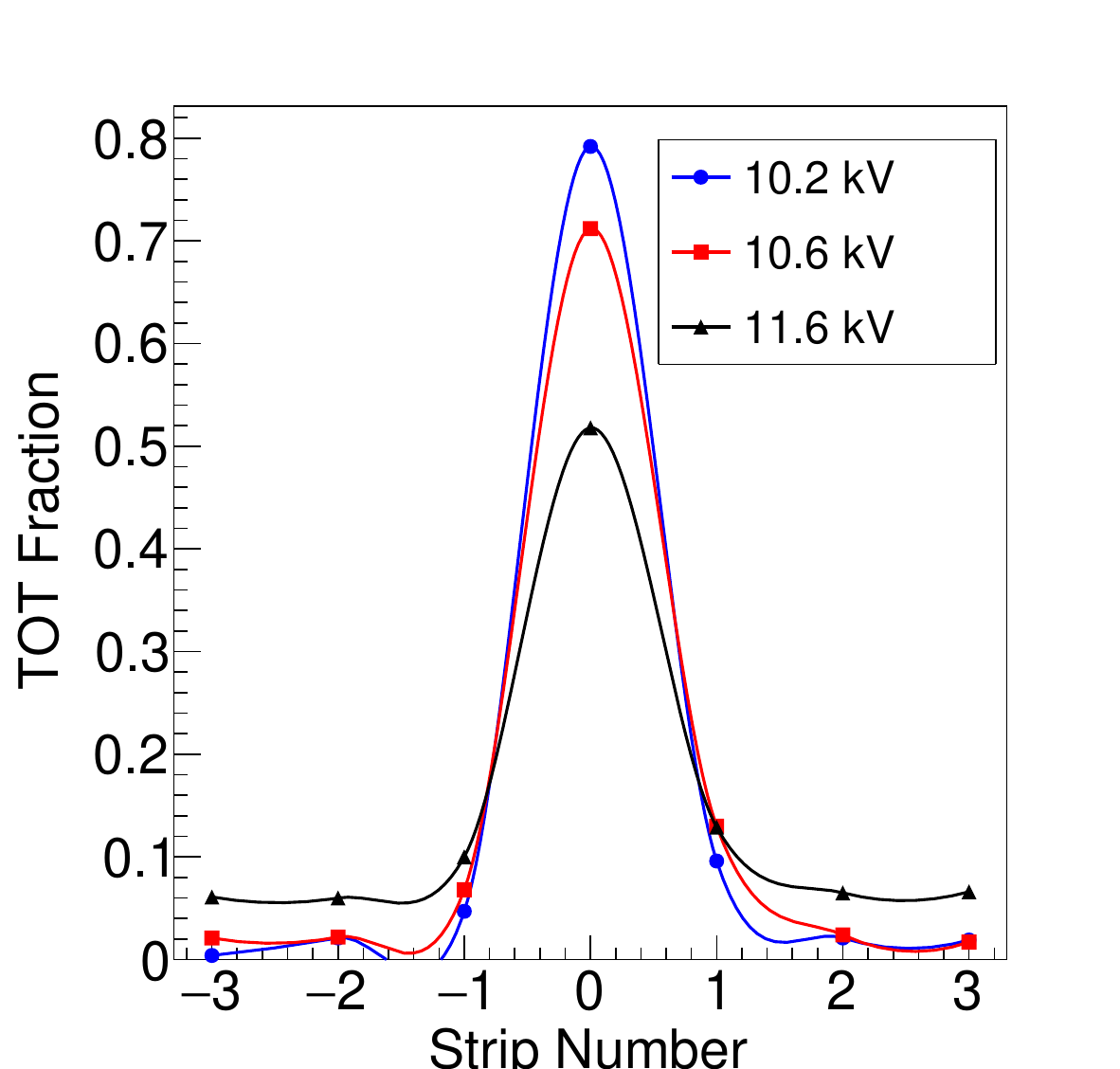}
	\caption{\label{TOTFrac} TOT fraction received by all the seven strips for three different working voltages}
\end{figure}
\section{TOT Measurement Results}
The TOT output of NINO has been found to increase with input pulse width linearly as shown in figure~\ref{widthTOT}. For a given signal with multiple strip hits, the maximum amount of charge is induced on the strip which covers the detector region where the particle has actually passed through. Therefore, the output of that strip would give the largest pulse width and hence the largest TOT among all the strips. This can provide the position information of the event. The TOT measurement has been carried out at three different working voltages (10.2 kV, 10.6 kV, 11.6 kV) to evaluate the tracking ability of this DAQ technique. The distribution of TOT received by the central strip for these voltages has been shown in figure~\ref{DiffVolt}. It may be noted that the larger tail of the distribution at 11.6 kV suggests transition to streamer mode. Moreover, the streamer signals have shown larger spread than the avalanche signals, as shown in figure~\ref{StripM}. It has been demonstrated that the multiplicity increases as the working voltage is raised. Finally, the fraction of TOT spread (ratio of TOT obtained from a given strip to total TOT obtained from all the strips) on the central strip and six neighbouring strips has been shown in figure~\ref{TOTFrac}. The standard deviations of the distributions has been found to increase with the rise in working voltage. At an avalanche working voltage (10.2 kV), the standard deviation of the spread of the TOT obtained is 4.5 mm with 1.2 cm strip pitch. This can be considered as a functional estimate of spatial resolution.
\section{Conclusion}
A novel scheme of read-out electronics for a RPC-based MST setup has been proposed using NINO ASIC in the front-end and an FPGA-based DAQ in the back-end. The TOT measurement of the NINO with a resolution of 2 ns has been achieved using a 500 MHz clock pulse derived using PLL on the on-board clock of FPGA. The preliminary tests have indicated that the present scheme can be utilised for retrieving position information with spatial resolution of the order of mm with the current design parameters of the RPC. The authors plan to fabricate a set of new read-out panel with narrower strips to achieve sub-millimeter spatial resolution. This can be useful for building an MST setup for material discrimination.

\acknowledgments
The authors are extremely grateful to the INO Collaboration for the support in electronics. The authors are thankful to Prof. Sandip Sarkar, MPGD lab members and Mr. Shaibal Saha for the experimental support and advice. The author, Sridhar Tripathy, acknowledges the support and cooperation extended by INSPIRE Division, DST, Govt. of India.

\end{document}